\documentclass[useAMS,usenatbib]{mn2e}
\input psfig.tex

\title[]{The history of mass assembly of faint red galaxies 
in 28 galaxy clusters since $\mathbf{z=1.3}$}
\author[S. Andreon]{S. Andreon,$^1$\thanks{stefano.andreon@brera.inaf.it}\\
$^1$INAF--Osservatorio Astronomico di Brera, Milano, Italy\\
}
\date{Accepted ... Received ...}
\pagerange{\pageref{firstpage}--\pageref{lastpage}}
\pubyear{2005}
\begin{document}
\maketitle

\label{firstpage}

\begin{abstract}
We measure the relative evolution of the
number of bright and faint (as faint as 0.05 $L^*$) red galaxies 
in a sample of 28 clusters, of which 16 are at $0.50\le z \le 1.27$,
all observed through a pair of filters bracketing the $4000$ \AA \ break
rest-frame. The abundance of red galaxies, relative to bright ones, is
constant over all the studied redshift range, $0<z<1.3$, and rules out 
a differential evolution between bright and faint red galaxies as
large as claimed in some past works. 
Faint red galaxies are largely assembled and in place
at $z=1.3$ and their abundance does not depend on cluster
mass, parametrized by velocity dispersion or X-ray luminosity.
Our analysis, with respect to previous one, samples a wider
redshift range,  
minimizes systematics and put a more attention to statistical 
issues, keeping at the same time a large number of clusters. 
\end{abstract}

\begin{keywords}  
Galaxies: evolution --- galaxies: clusters: general --- 
Galaxies: luminosity function, mass function --- Galaxies: formation 
\end{keywords}

\section{Introduction}

The evolution of faint red galaxies in clusters is a highly debated topic  
for two reasons: different observers
have claimed controversial results, and clusters of galaxies are often
claimed to be interesting laboratories where studying the effect of
the environment. Red galaxies, in particular, have
different assembly histories in halos of different masses, 
yet observationally the detection of a
environmental dependence of their properties escapes a detection.
For example, differences between cluster and field
fundamental planes are small, if any (Pahre et al. 1998), so
small that the Coma cluster fundamental plane (Jorgensen et al. 1996) 
is routinely used as zero-redshift reference for studying
the evolution of field galaxies, and so small
that previously claimed differences are probably due
to having overlooked the difficulty of
the statistical analysis (van Dokkum \& van der Marel 2007). Similarly,
the colour of the red sequence seems not to depend on clustercentric
distance (Pimbblet et al. 2002; Andreon 2003) or galaxy number density (Hogg et al. 2004, Cool
et al. 2006).

The red colour, by which red galaxies are defined and selected,
induces a selection effect: at
every redshift only galaxies whose stellar populations are red
(i.e. old, modulo dust, of no interest here)
enter the sample. It is not a surprise, then, to find
old-selected populations to be old. A different question is
whether galaxies that have an old stellar population were fully
assembled at early or late times. Answering this question requires 
a measurement of the abundance of red galaxies as a function
of look-back time. For clusters, there is a further complication:
clusters have different richnesses, jeopardizing any 
look-back time trend if the 
richness dependence is not
factored out. It is easy, furthermore, to qualitatively claim that 
the red sequence is built later (i.e. a lower redshift) in
poor environments than it is in dense 
environments, but this might just be
do to signal to noise issues, because in poorer environments
the red population is a minority one, and its contrast
with respect to other populations (e.g. background) noisier. 
A sound statistical assessment of the abundance of faint
red galaxies is therefore compelling.

\begin{table}
\caption{The ACS $z\ge0.5$ cluster and control field samples}
\begin{tabular}{l r c c l l}
\hline
Name & z & $N^{1}$ &\multispan{2}{\hfill Filters \hfill} \\ 
     &   &         & blue & red \\
\hline
Lynx W            & 1.27 &  3	 &  F775W &   F850LP \\   
Lynx E            & 1.26 &  3	 &  F775W &   F850LP \\    
RDCS J1252-2927   & 1.23 &  4	 &  F775W &   F850LP \\    
RDCS J0910+5422   & 1.11 &  1	 &  F775W &   F850LP \\    
GHO 1602+4329	  & 0.92 &  1	 &  F606W &   F814W  \\    
GHO 1602+4312	  & 0.90 &  1	 &  F606W &   F814W  \\    
1WGA J1226.9+3332 & 0.89 &  6	 &  F606W &   F814W  \\    
MACS J0744.8+3927 & 0.70 &  1	 &  F555W &   F814W \\
MACS J2129.4-0741 & 0.59 &  1	 &  F555W &   F814W \\        
MACS J0717.5+3745 & 0.55 &  1	 &  F555W &   F814W \\        
MACS J1423.8+2404 & 0.54 &  1	 &  F555W &   F814W \\        
MACS J1149.5+2223 & 0.54 &  1	 &  F555W &   F814W \\        
MACS J0911.2+1746 & 0.50 &  1	 &  F555W &   F814W \\ 
MACS J2214.9-1359 & 0.50 &  1	 &  F555W &   F814W \\ 
MACS J0257.1-2325 & 0.50 &  1	 &  F555W &   F814W \\ 
 \\ 	  
CT344	   &  & 1 &  F606W &   F814W  \\    
B0455	   &  & 1 &  F555W &   F814W  \\    
GOODS+PAN  &  & $\sim30$  &  F775W &   F850LP   \\ 
\hline                                                      
\end{tabular} \hfill \break
\footnotesize{$^1$ number of ACS field of view per filter. \hfill\break
All clusters have coordinates and redshift listed 
in NED, except for MACS clusters, listed
in Ebeling et al. (2007).  MACS clusters have been also studied by Stott
et al. (2007).
\hfill\break}
\end{table}

Usually,
the richness dependence of the  abundance of faint red galaxies 
is removed by normalizing it  
to the number of bright red galaxies, i.e.
by computing the faint-to-luminous ratio, or any related quantity, like
the faint-end slope $\alpha$ of the luminosity function.
The analysis of the faint-to-luminous ratio, performed by Stott et 
al. (2007), or its reciprocal,
the luminous-to-faint ratio by
De Lucia et al. (2007), both suggest an evolution of the relative
abundance of faint red galaxies, in the sense that at high
redshift there is a deficit of faint red galaxies per unit
bright galaxy. 
On the other end, Andreon (2006a) suggests no deficit of red galaxies, 
using a very small cluster sample, and Andreon et al. (2006)
discard a considerable build up of the red sequence on the basis
of fossil evidence. Evidences presented in earlier works have 
been discussed in the mentioned papers and references therein.

In this paper, we aim to understand if the colour--magnitude
relation has been build up at early or late times, by studying
many galaxy clusters at several look-back times.

Throughout this paper we assume $\Omega_M=0.3$, $\Omega_\Lambda=0.7$ 
and $H_0=70$ km s$^{-1}$ Mpc$^{-1}$. All
results of our stochastical computations are quoted in
the form $x\pm\sigma$ where $x$ is the posterior
mean and $\sigma$ is the posterior standard deviation.

\section{Data \& data reduction}

 This work makes use of archive data. The selection criteria used 
for the inclusion in our sample are the following: a) all observations
must include a pair of filters bracketing the 4000 \AA \
break in the cluster rest-frame; b) control field observations 
with identical conditions (same
telescope, instrument, filters, dept and seeing) as cluster observations
must be available; c) observations
had to be deep enough to measure the faint end slope of the
luminosity function; d) clusters had to be spectroscopically 
confirmed; e) data should be public available at the start of
the work.

Our sample is formed by three sets: a) 15 high redshift clusters
observed with the Wide Field Camera of the Advanced Camera
for Surveys (hereafter ACS, Ford et al. 1998, 2002) of
{\it Hubble Space Telescope} (HST, hereafter); b) two low redshift
clusters observed by the Sloan Digital Sky Survey (SDSS); and c)
two luminosity functions from literature, one for a $z\sim0.25$
cluster sample and one for one more high redshift cluster
observed by HST but with the Wide Field Planetary Camera 2.

Table 1 lists the ACS sample,
formed by 15 
clusters at $0.5\le z \le 1.27$. More than 150 HST orbits, devoted
to clusters, have been
reduced and analized for this paper.
As we need to statistically  discriminate 
against fore- and back-ground interlopers, Table 1 also lists
the adopted control fields. A control
field matching the filter pair used for clusters is available
for all targets. 

In order to provide a local ($z \sim 0$) reference, 
we use SDSS $u,g$ data of two nearby clusters:
Abell 1656 (A1656 hereafter,  i.e. Coma, $z=0.023$) and 
Abell 2199 (A2199 hereafter, $z=0.030$). Given the large SDSS sky
coverage, the control 
field for our nearby clusters is taken all around them.

Finally, the luminosity function of ten $z\sim0.25$ clusters, 
observed in $B$ and $I$ by Smail et al. (1998), and of MS1054 at $z\sim0.8$, observed
with HST Wide Field Planetary Camera 2 in F606W and F814W
and presented in Andreon (2006) have been 
taken from the literature.  These LFs are fully homogeneous to those
computed in this work.

Figure 1 shows that 
all clusters have a pair of filters bracketing the 4000 \AA \
break.

\begin{figure}
\psfig{figure=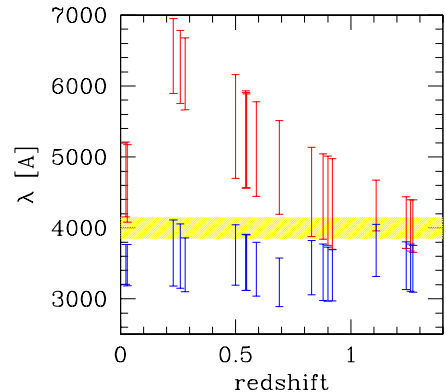,width=6truecm,clip=}
\caption[h]{Rest-frame $\lambda$ sampling of the adopted filters for the clusters
studied in this work. The shaded (yellow) band marks the
4000 \AA \ break. All clusters have been observed in a pair of filters
bracketing the 4000 \AA \ break.  
The sample is formed by 28 clusters, some
of which have very similar redshifts and do not show up individually
in the Figure.
}
\end{figure}

\begin{figure}
\psfig{figure=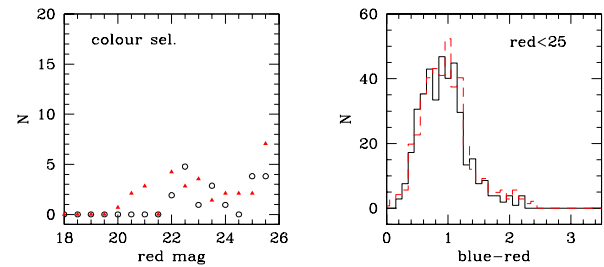,width=8truecm,clip=}
\caption[h]{Minor role of non-Poisson
variance. The left panel shows galaxy counts of red galaxies 
in two widely different sky direction, whereas the right panel shows their
colour distribution. Differences between sky directions are within
Poisson fluctuations.
}
\end{figure}

\begin{figure*}
\centerline{%
\psfig{figure=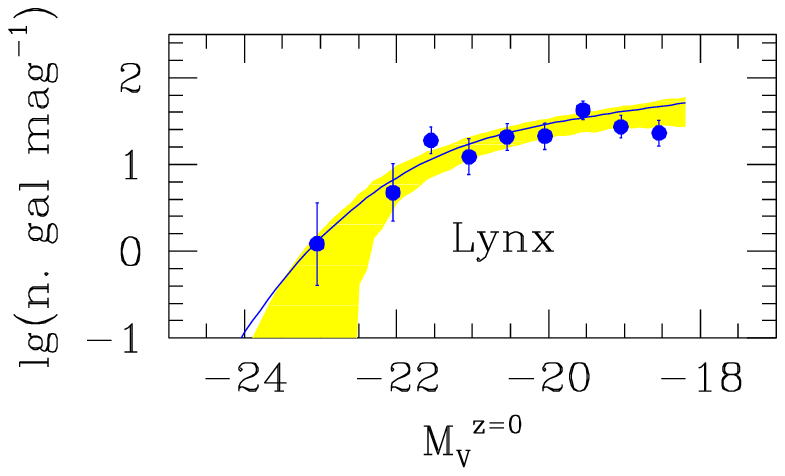,width=6truecm,clip=} \hfill
\psfig{figure=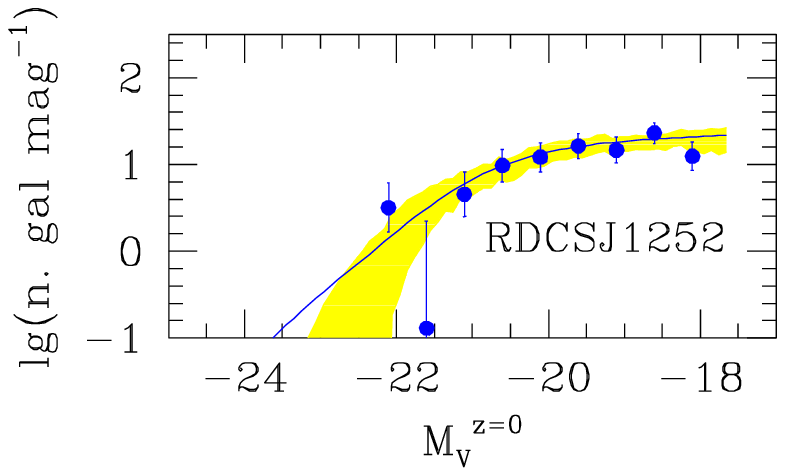,width=6truecm,clip=} \hfill
\psfig{figure=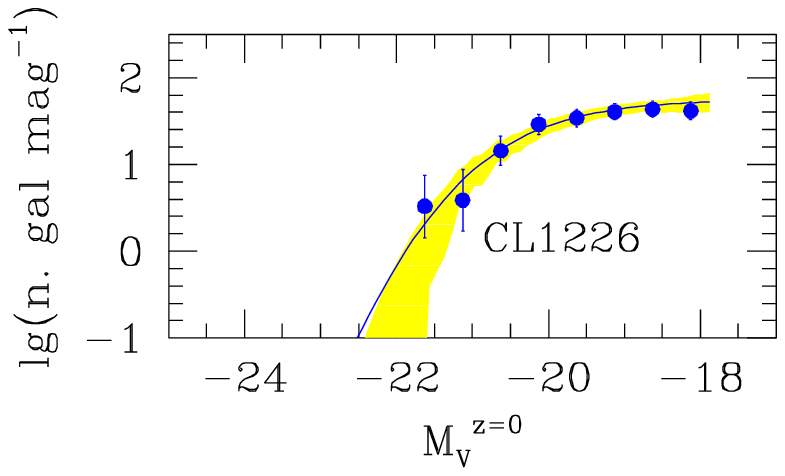,width=6truecm,clip=} \hfill
}
\centerline{%
\psfig{figure=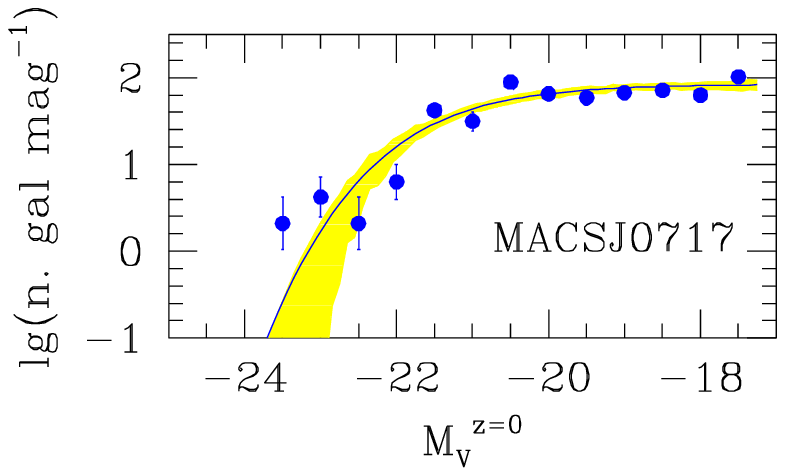,width=6truecm,clip=} 
\psfig{figure=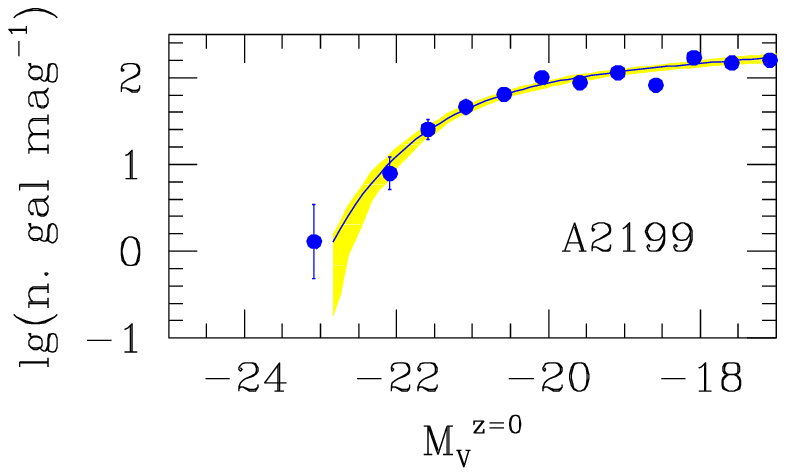,width=6truecm,clip=} 
\psfig{figure=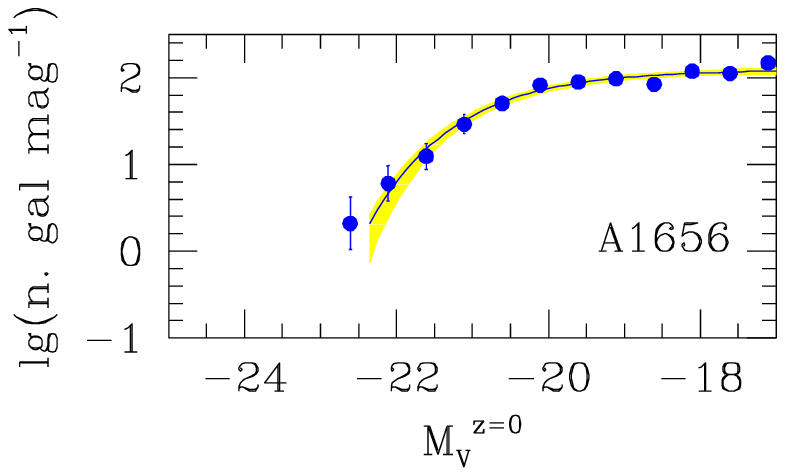,width=6truecm,clip=}
}
\caption[h]{Luminosity function of red galaxies in some (to 
save space) of our 
clusters. The solid curve and the shaded region mark the LF and its
(highest posterior) 68 \% error region as computed with Bayesian methods.
Points and error bars mark simply-derived LFs, computed as difference of
counts in the cluster and reference lines of sight, using
``two $\sigma$ from the colour-magnitude relation'' as definition of red.
The top-left panel refers to the two Lynx clusters stacked together. 
}
\end{figure*}

The raw ACS data listed in Table 1 
were processed through the standard CALACS pipeline (Hack 1999)
at STScI. This includes overscan, bias, and dark subtraction, as well as
flat-fielding. Image
combination has been done with the multidrizzle software (Koekemoer et al. 2002). 
The data quality arrays enable
masking of known hot pixels and bad columns, while cosmic rays and other
anomalies are rejected through the iterative drizzle/blot technique. Sources are
detected using SExtractor (Bertin \& Arnout 1996), making use of weight maps
produced by Multidrizzle.
Star/galaxy separation is performed by using the stellarity index given by
SExtractor.
HST images are calibrated in the Vega system, using the zero points provided in the
HST data handbook. 
Completeness is computed as in Garilli, Maccagni \& Andreon (1999), from
the brightest luminosity of the detected objects of faintest surface
brightness. Only data brighter than the completeness magnitude are kept.

All science (i.e. cluster) and two of the control fields, CT344 and BO0455, have
been combined (and catalog built) by ourself, while the remaining control field, 
GOODS+PAN, has been generously given to us by D. Macchetto. 
These
images come from the same telescope, instrument and filters and have been
processed with the same software as science data (i.e. CALACS, Multidrizzle and
SExtractor), but have been combined by someone else (than the author). 
By reducing by ourself part ot the GOODS+PAN data, we checked
that their and our reductions are indistinguishable.

The left panel of Figure 2 shows galaxy counts of red galaxies, where
`red' is taken to mimic our later selection,
for two widely different sky
directions: CT344 and a field in Benitez et al. (2003). 
 The right panel shows the colour
distribution in the two directions.  
Differences between the two sky directions are comparable to
Poisson errors on the average value. Therefore,  for
areas, magnitudes and colours of interest in this paper, non-Poisson
fluctuations of galaxy counts can be neglected.

For the nearby cluster sample, catalogs have been extracted 
from the SDSS 5th data release 
(Adelman-McCarthy et al., 2007), which have been produced by the 
SDSS pipeline and are not calibrated in the Vega system.
We checked that synthetic $U,V$ and $U-V$ computed from
$u,g$ SDSS photometry is indistinguishable from observed $U, V$
and $U-V$ photometry for red galaxies in A1656 cluster direction,
taken from Terlevich
et al. (2001), and derived with traditional
techniques (stare exposures, calibration in the Vega system, and catalogs
built with SExtractor).

\section{Faint-to-luminous ratio of red galaxies}

We modelled the distribution of galaxies in the red
sequence as Gauss-distributed in colour at every  magnitude and
Schechter (1976) distributed in magnitude. The mean colour of the
Gauss varies 
linearly with magnitude, because the color-magnitude
relation is linear. Furthermore, we allow a broadening of the
colour-magnitude relation due to both
photometric errors and an intrinsic scatter.  
As explained in appendix A in Andreon (2006a), 
with Bayesian methods we solved at once
for all parameters (colour-magnitude slope, intercept and intrinsic
scatter, characteristic magnitude $M^*$, faint-end slope $\alpha$,
and normalization of the Schechter, background parameters), 
hence fully accounting for
the background (including uncertainty, variance and covariance
with all parameters). 

Our definition of red is ``galaxies under the Gauss centered 
on the red sequence", similar to some SDSS works
(e.g. Balogh et al. 2004; Ball et al. 2006). Previous studies 
(Andreon et al. 2006, De Lucia et al. 2007) have shown that the
precise definition of `red` has a negligible impact on the results. 
We have checked it for our own sample and our definition,
by adopting a simpler definition of red (within $2 \sigma$ from
the red sequence, plotted as dots in Fig 3).

The luminous-to-faint ratio is computed as the ratio of the number of
galaxies on the red sequence in appropriate absolute magnitude ranges.
The number of galaxies in a given range is, by
definition, the integral of the luminosity function 
over the concerned range. The range definitions are taken from
De Lucia et al. (2007): $M_V<-20$ mag and 
$-20<M_V<-18.2$ mag. 
Magnitudes are passively evolving, modelled with a simple
stellar population of solar metallicity, Salpeter IMF, from Bruzual \&
Charlot (2003), as in De Lucia et al. (2007). As a sanity check, the same model has
been checked to reproduce the colour of the red sequence at 
$M_V = -20.0$ mag for all our clusters. 

From now on, the two Lynx clusters are stacked together 
to improve the signal to noise.
LFs are computed for galaxies within a cluster-centric radius
listed in Table 2. The considered region has
been chosen as a compromise between sampling a large portion of
the cluster and
not including a too large contribution from background galaxies.
MACS clusters are larger that the instrument field of view, and
therefore we choose the largest radius that fit in the fully exposed
part of the image, consistently with the choices of
Andreon (2006) and Smail et al. (1998), whose LF are included
in the present work, as mentioned.

The potential dependency of the LF slope on
the considered cluster portion has a small impact on our
study, because
we explicitely allows the observed value of the slope to
scatter around to its true value by more than its uncertainty.
In fact, in the next section we  
we allow an intrinsic scatter in our model: see eq. 1. This argument
is developed further in Sec 4.2.

\section{Results}

Figure 3 shows the luminosity function of a sub-set (to save space)
of studied clusters. 

Figure 4 shows the luminous-to-faint ratio, $L/F$, as a function of redshift
for clusters with LF measured in this paper (solid points).
Our data are in agreement with De Lucia et al. (2007) data 
(open points), but our wider redshift coverage suggests a shallower
trend than the one hinted in De Lucia et al. (2007) from their
data points.

In this work we refrain to perform inferences using $L/F$ or its
reciprocal, $F/L$,
for reasons detailed in sec 4.4, mainly of statistical nature.
The use of the faint end slope, $\alpha$, is a 
measure of the faint-to-luminous ratio, it is easier to
deal with from a statistical point of view, and has the further 
advantage that it uses all the
data available, including data fainter than $-18.2$ mag
that would be otherwise wasted using $L/F$.

Figure 5 shows the slope, $\alpha$, as a function of redshift for the whole
cluster sample, i.e. for 28 clusters, of which 16 at $z\ge 0.5$.
Marginalization accounts for the 
known correlation between parameters (e.g. $M^*$ and $\alpha$). For example,
the large error of some data points is due to the fact that many
($M^*$,$\alpha$) pairs fit almost equally well the data and thus a large
range of $\alpha$ values is acceptable. $\alpha$ errors also account
for differences in the galaxy background counts in the cluster and
control field lines of sight, because, as mentioned, we ``solve" for all
parameters at once (technically, we marginalize over other parameters).
Table 2 lists the $\alpha$ and $L/F$ values found.

\begin{figure}
\psfig{figure=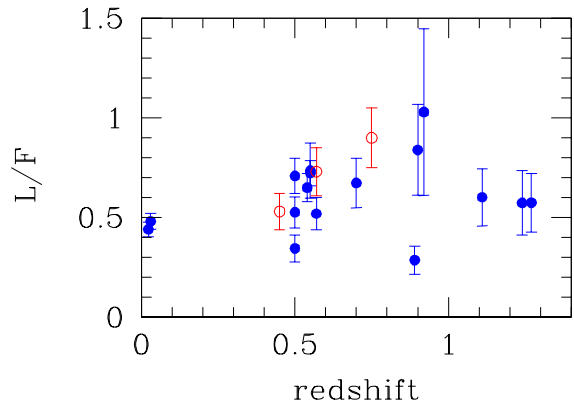,width=8truecm,clip=}
\caption[h]{Relative abundance of faint and bright red galaxies, as 
parametrized by the luminous-to-faint ratio,
for clusters with LF measured in this paper (solid dots) and in 
De Lucia et al. (2007, open points). Although in agreement,
our data indicate a shallower trend with redshift than De Lucia et al. 
(2007) data points. Two points at $z=0.55$ fall one on the top of 
the other.
}
\end{figure}

\begin{figure}
\psfig{figure=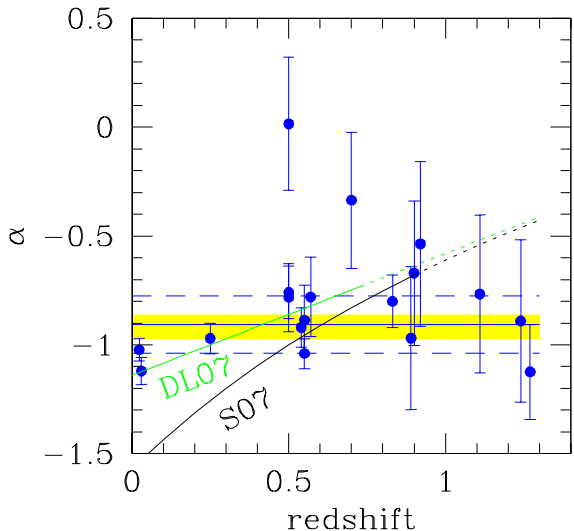,width=8truecm,clip=}
\caption[h]{Relative abundance of faint and bright red galaxies, 
as parametrized by the faint slope $\alpha$ of the 
cluster luminosity function for the whole sample of 28 clusters
studied here. The redshift dependence of
the relative abundance of bright red galaxies is small, if any.
The point at $z=0.25$ ($z=1.27$) is the average of 10 (2) clusters.
The shaded
(yellow) region shows the (highest posterior density) 68 \% error
region. 
The dashed lines delimit the $\pm 1 \sigma$ intrinsic (i.e
not accounted for measurement error) scatter.
The trends proposed by Stott et al. (2007) and we fitted on De Lucia
et al. (2007) data are also marked with solid lines (labelled by ``S07" and 
``DL07", respectively)
up the the largest studied redshift by them, and marked
with dotted lines afterward. 
}
\end{figure}

The data are in agreement with the lack of a deficit of
faint red galaxies suggested by Andreon (2006) on the basis
of a very small sample of clusters and
reject some trends suggested in previous works.
Let's consider: a) our maximum likelihood fit of the $L/F$ data
points in Fig. 9 of De Lucia et al. (2007) and b) the Stott et al. (2007)  
$F/L$ vs $z$ fit. The two fits have been transformed in $\alpha$ vs $z$ 
trends using the $L/F$, $F/L$ and $\alpha$ definitions. 
Figure 5 shows that at low and intermediate redshift
the De Lucia et al. (2007) trend, marked with ``DL07", is compatible
with our data. However, a constant, i.e. 
a more economical model having one degree
of freedom less, also well describes our data (and also theirs, see Fig. 5)
over the common redshift range ($z<0.8$) and, actually, also above. 
Furthermore, neither De Lucia et al. (2007) nor our data
request a more complex model than a constant plus an intrinsic
scatter. The computation of the Bayes factor shows that
the De Lucia et al. (2007) trend is disfavored, with respect to
`no trend at all' by our data with odds
14:1, i.e. there is moderate evidence against an increase of the 
luminous-to-faint ratio as large as pointed out by De Lucia
et al. (2007). We refrain, therefore, from fitting a 
more complex model, and we adopt a constant model. 
Figure 5 also shows that the Stott et al. (2007) fit, marked with ''S07" in 
the Figure, nicely reproduces the observed values in the reduced
redshift range, $0.5 \le z \le 0.6$, where we share clusters and 
HST data with them, but disagrees outside it, in particular at low redshifts. 
Furthermore,
in the local universe, the Stott et al. (2007) fit 
and data also disagrees with De Lucia et al. (2007)
data and trend. Our data clearly discard the trend proposed by Stott et al. (2007).

Using Bayesian methods (D'Agostini 2003, 2005) {and uniform priors}
we 'fitted' the data point with a constant, accounting for
errors and allowing an intrinsic (i.e. not accounted for errors) 
Gaussian scatter, $\mathcal{N}(0,\sigma_{intr})$. We found: 

\begin{equation}
\alpha (z) =  -0.91 \pm0.06 +\mathcal{N}(0,0.13\pm0.06) 
\end{equation}

displayed in Fig 5. Using our own data alone, i.e. ignoring
the Smail et al. (1998) $z=0.25$ composite cluster, we found an 
indistinguishable result:

\begin{equation}
\alpha (z) =  -0.89 \pm0.06 +\mathcal{N}(0,0.16\pm0.06) 
\end{equation}

\begin{figure}
\psfig{figure=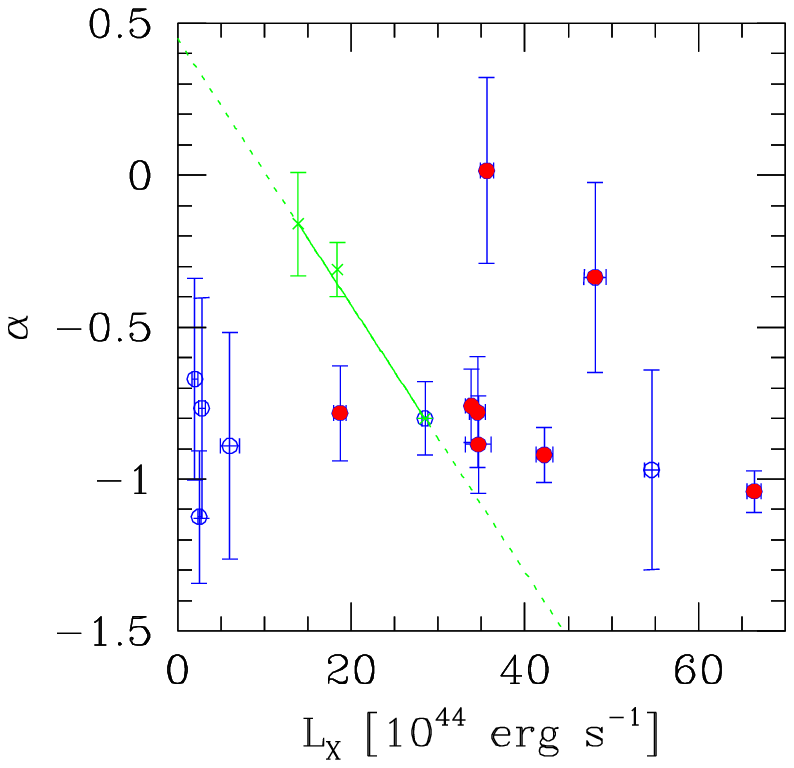,width=6truecm,clip=}
\psfig{figure=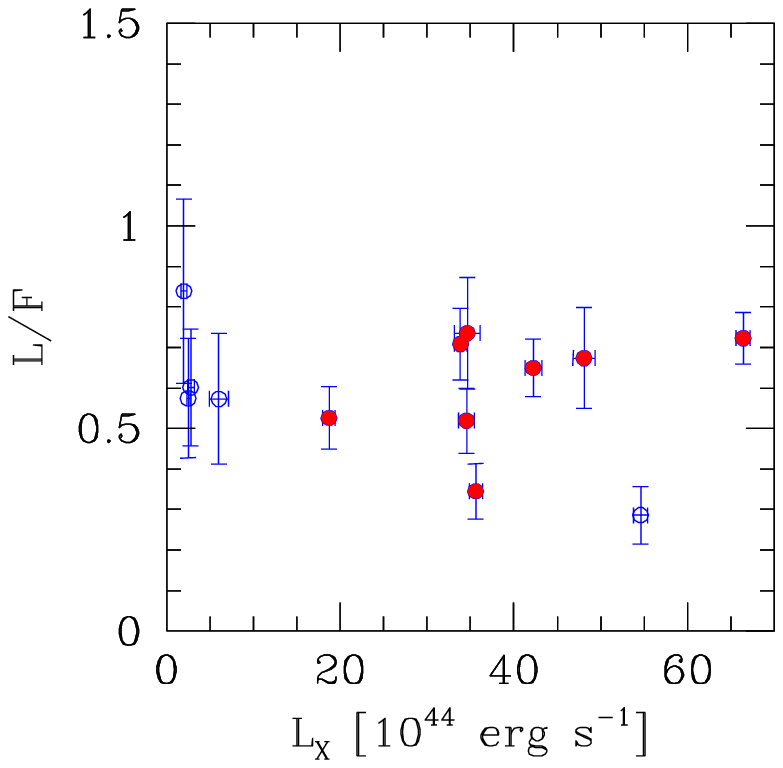,width=6truecm,clip=}
\caption[h]{Cluster mass dependency, as parametrized by X-ray bolometric
luminosity, of the relative abundance of faint red galaxies, as 
parametrized by the $\alpha$ (top panel) or the luminous-to-faint ratio
(bottom panel), for clusters with $z\ge 0.5$. Clusters in the 
$0.5\le z \le 0.70$ range  are marked with a solid (red) dot. The three
crosses connected by a solid line in top panel mark clusters used by
Koyama et al. (2007) to suggest a mass-dependent trend in the relative
abundance of faint red galaxies. 
}
\end{figure}

\begin{figure}
\psfig{figure=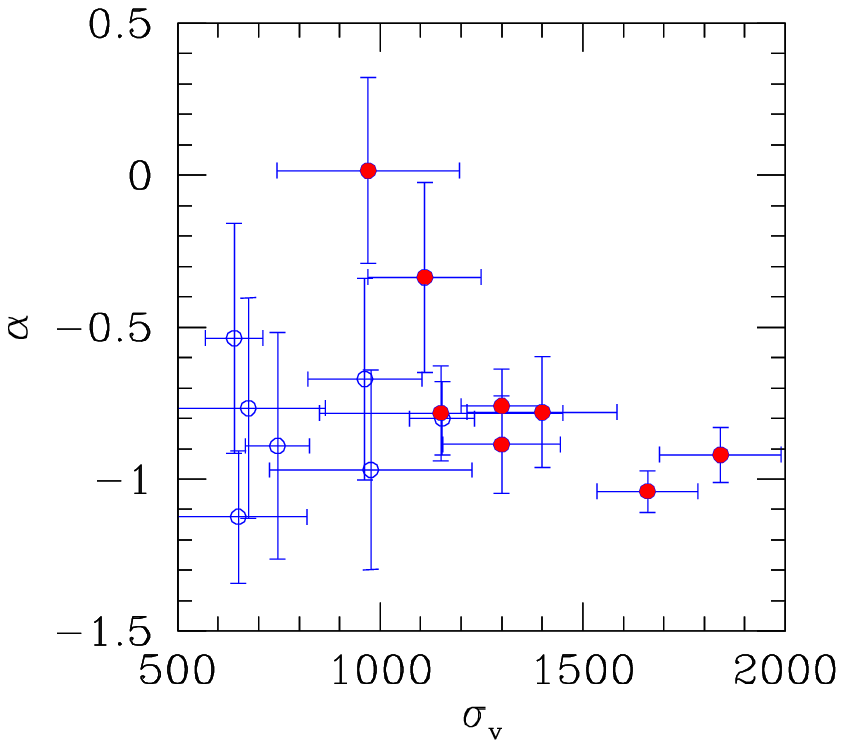,width=6truecm,clip=}
\psfig{figure=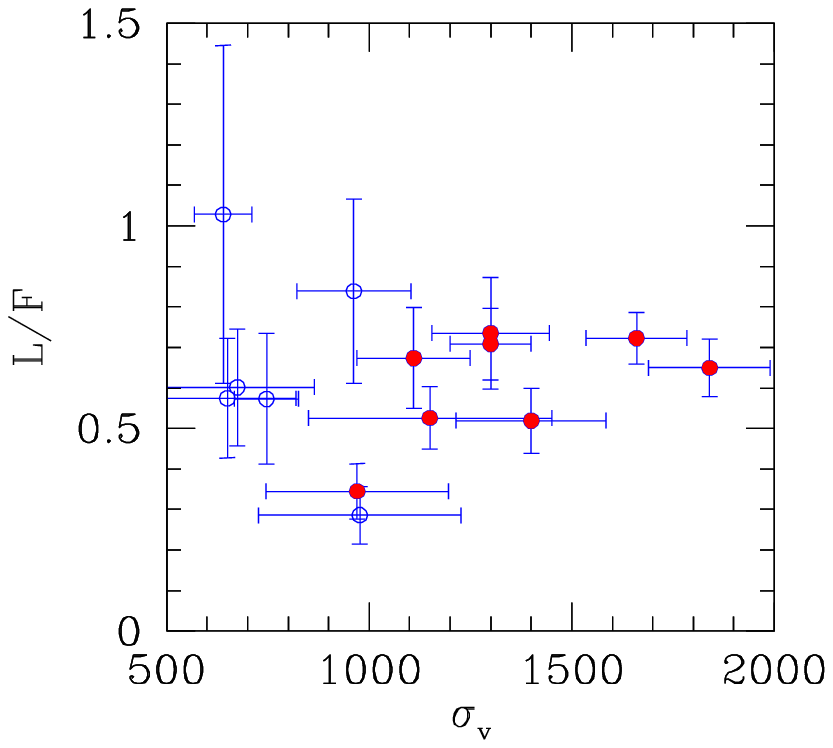,width=6truecm,clip=}
\caption[h]{As Figure 6, but using velocity dispersion as
cluster mass estimator.
}
\end{figure}

\subsection{Richness dependency}

Koyama et al. (2007), on the basis of three clusters at $z\sim 0.8$, suggested
that the relative abundance of faint red galaxies is dependent on cluster
richness or  mass (actually, X-ray luminosity in their work), in the sense that
poorer systems show stronger deficits. Fig. 6 and 7 plot two deficit estimators,
the slope $\alpha$ and the $L/F$ ratio, {\it vs} two mass estimators, X-ray
luminosity and velocity dispersion, for clusters at $z\ge 0.5$. X-ray
luminosities are taken from Ettori et al. (2004), Lubin et al. (2004) and
Ebeling et al. (2007).  They comes from Chandra or XMM
pointed observations and have, typically, errors of 10 \% or less.
MACS luminosities, in the 0.1-2.4 keV band are converted in bolometric
assuming a thermal bremsstrahlung spectrum with the measured temperature.
Velocity dispersions are taken from Ebeling et al. (2007), Stanford et al.
(2001) and Maughan et al. (2004). Solid dots emphasize clusters in the reduced
redshift range $0.50 \le z \le 0.70$, to limit the (negligible, see previous
section) effect of evolution. There is no obvious trend between cluster mass and
the relative abundance of faint red galaxies. (Green) crosses, we connected by
a solid line in the top panel of Fig 6, show the three clusters studied by
Koyama et al. (2007), one of which is MS1054.4-0321, the latter taken from
Andreon (2006). The slope $\alpha$ is derived by us from their luminous over
faint ratio assuming a Schechter function. The figure shows that the observed
slopes $\alpha$ are plausible, since two other clusters in our sample show
similar values of the relative abundance of faint red galaxies. However, the
trend suggested by these three points (slanted line) is clearly too steep, and
obviously ruled out  by our data.  Finally, we note that RXJ0152.7-1357 point,
i.e. the middle point of the three plotted, has been put in Koyama et al.
(2007)  and in our plot at the {\it sum} luminosity of the two sub-clumps that
form the cluster, not at the {\it mean} $L/F$ value. Would the RXJ0152.7-1357
point be put at the average x-ray luminosity of the two clumps, the one
typically experienced by galaxies in this clusters and consistently with the
choice of quoting a mean $L/F$, the three $L/F$ points would no longer show any
monotonic trend with X-ray luminosity. 

In conclusion, the
abundance of faint red galaxies does not considerably depend on cluster mass
(in the range sampled by data, of course) with the Koyama et al. (2007) trend
largely based on a sample of inadequate size, given the 
large intrinsic scatter, a possibility also mentioned by these authors.

\subsection{Radius- and scatter- related effects and the advantages 
of allowing an intrinsic scatter}

Clusters have no sharp boundaries. In order to understand the potential
effect of the choice of the studied cluster portion, we compute $\alpha$ and
$L/F$ of A1656 cluster within two clustercentric radii: 0.2 and 0.8 deg. 
A1656 cluster has been chosen because it has the 
best determined values of $\alpha$ and $L/F$ among all
our clusters. For A1656, the relative abundance of faint red galaxies, 
as parametrized by
these quantities, is the same within the two radii. Although this
test is reassuring, we cannot generalize from a single example.

Our
model, eq. 1, explicitly allows an intrinsic variance in the relative abundance of
faint red galaxies, due for example to the mentioned clustercentric dependence,
cluster-to-cluster or other possible (unknown, for the time being) systematics.
These terms are thus a source of scatter, not different from a random number
added to each measurement. We model such random process, whatever its physical 
nature is related to clustercentric distance, to cluster-to-cluster variance
or whatsoever unidentified reason, with a normal
(Gaussian) of unknown variance (eq. 1), for
lack of evidence toward any more complex model. In passing, in absence of more
information, the Gaussian is the maximum entropy choice among all real-valued
distributions with specified mean and standard deviation (e.g. Sivia 2006). The
sum rule of probability states that in order to proceed with the inference, we
need to marginalize over this (nuisance) parameter. 
The intrinsic scatter
parameter, and its consequent marginalization, offers protection against
claiming a trend when just too few data are available: marginalization
spreads the probability of a (redshift or mass) trend over a
large range of slope values, i.e. quantifies the
researcher good sense that when just a few data points are available
and an intrinsic scatter is there, one should be
prudent in claiming the existence of a trend.
Assuming a single value for the
intrinsic scatter, as other authors sometime implicitly take when looking
for a trend, artificially collapses the error ellipse along this axis 
and leads to determinations with overly-optimistic confidence.

Therefore, since the model allows an intrinsic scatter,  
the analysis of the redshift depencence of the relative aboundance
of faint red galaxies is correct even if:
a) a clustercentric dependence of the
the relative abundance of faint red galaxies exists, provided that we are not 
sampling increasing
smaller cluster portions as the redshift increases or decreases, or b) the relative
abundance of faint red galaxies differs from cluster to cluster.  Both
cases are just source of scatter and our model account for them.

As mentioned, we found a non-zero
intrinsic  (i.e. not accounted for measurement errors)  scatter
($\sigma_{intr}=0.13 \pm 0.06$),  quantifying past claims of a heterogeneity in
the relative abundance of faint red galaxies. Inspection of the colour-magnitude
relations of outliers in Fig 5, e.g. MACS J0257.1-2325, confirms that these
clusters have an underpopulated red sequence at faint magnitude or an
overpopulated one at bright ones.  

Therefore, in the quest of a build-up of the red sequence,
an intrinsic scatter must be allowed, in order
not to overweigh 'outlier' clusters, and not to overstate
the precision and the statistical significance of the found (redshift, mass
or whatever) trend.
We note that the existence of an intrinsic scatter has been claimed
in previous works (De Lucia et al. 2007, Stott et al. 2007)  
 but ignored (Stott
et al. 2007) or not rigorously accounted for (De Lucia et al. 2007) when
establishing the veracity of the claimed redshift trend.

The existence of an intrinsic scatter testifies that: a) there is a 
yet to be identified physical mechanism that affects the relative 
abundance of faint red galaxies, and b) current data are of adequate
quality to perform such measurement, i.e. that the topic deserves 
further investigation.

\subsection{Joining high- and low- redshift information}

We emphasize that our (past) knowledge about individual nearby 
clusters tell us that at least some galaxies on the red sequence
have a spiral morphology (Butcher \& Oemler 1984;
Oemler 1992; see Fig. 3a in
Andreon et al. 1996 or Fig. 4 in Terlevich et al. 2001 for A1656
galaxies). Their spiral arms
testify that, in the past, these galaxies were forming
stars, i.e. were blue, and therefore were not on the red sequence.
Furthermore, at least for A1656 (Coma), red spirals have lower surface 
brightness than blue spirals (Andreon 1996), as
expected if the former are the descendent of the latter.
Since, on average, spirals are fainter than early-type galaxies
(e.g. Bingelli, Sandage, Tammann 1988 for Virgo, Andreon 1996
for A1656), we expect that the abundance of faint red galaxies
grows somewhat with time, just because of the evolving colour
(toward the red) of some spirals. However, it cannot grow too much,
otherwise it would bend the colour-magnitude and inflate its
scatter. The argument is the usual one (e.g. Bower, Lucey \& Ellis
1992): a) a heterogeneity in the star formation history leads
to a heterogenous population in colour (unless something else
coordinately conspires to keep the colour scatter small); and b)
a delayed stop of the last star formation episode 
delays the arrival of a galaxy on the colour-magnitude relation,
bending it (or increasing the colour scatter if there 
is an un-delayed population).
In Abell 1185 the colour-magnitude relation is
linear and the scatter in colour is small ($0.04$ mag) down
to $M^*+8$ (Andreon et al. 2006). In A1656, the scatter stays
constant to low levels ($0.05$ mag) down to $M^*+4$ (Eisenstein et al.
2007). Therefore, fossil evidence points toward a small, but not null, 
differential build up of red sequence galaxies. 

Theory (De Lucia et al. 2007) 
shows that a model in which star formation histories of blue
galaxies are truncated produces an important change in the
luminous-to-faint ratio, larger than allowed by De Lucia et al. (2007) data, 
producing too many faint galaxies by a factor two (but note that these
authors consider
it as 'approximatively consistent' with their data). Direct measurement
of the abundance of red faint galaxies over redshift (this paper)
indicates a shallower trend with redshift 
than suggested by theory at a point that data
are consistent with no trend at all. 

Therefore, data at cosmological redshifts and the tightness of
the colour-magnitude relation at low redshift strongly
argue against a scenario where many blue galaxies transform themselves
in faint red galaxies, whereas the presence of
some spiral galaxies on the red sequence in nearby clusters suggests
a redshift trend in the relative abundance of faint red galaxies
should be observed in sufficiently large cluster samples, although
none is clearly revealed in the present one.

\begin{table}
\caption{Extraction radius $r$ in arcmin, faint end slopes $\alpha$ 
and luminous-to-faint ratios $L/F$ }
\begin{tabular}{l r c c}
\hline
Cluster name  & $r$ & $\alpha$ & $L/F$ \\
\hline
Lynx E+W    & 1.0 & $-1.12\pm0.22$ &  $0.57\pm0.15$\\   
RDCS J1252-2927   & 1.0 & $-0.89\pm0.37$ &  $0.57\pm0.16$\\    
RDCS J0910+5422   & 1.0 & $-0.77\pm0.36$ &  $0.60\pm0.14$\\    
GHO 1602+4329	 & 1.0 & $-0.54\pm0.38$ &  $1.03\pm0.42$\\	 
GHO 1602+4312	 & 1.0 &  $-0.67\pm0.33$ &  $0.84\pm0.23$\\	 
1WGA J1226.9+3332 (CL1226) & 1.5 &  $-0.97\pm0.33$ &  $0.29\pm0.07$\\    
MACS J0744.8+3927   & 1.4 &  $-0.34\pm0.31$ &  $0.67\pm0.12$\\
MACS J2129.4-0741   & 1.4 & $-0.78\pm0.18$ &  $0.52\pm0.08$\\	
MACS J0717.5+3745   & 1.4 & $-1.04\pm0.07$ &  $0.72\pm0.06$\\	
MACS J1423.8+2404   & 1.3 &  $-0.89\pm0.16$ &  $0.74\pm0.14$\\	
MACS J1149.5+2223   & 1.2 & $-0.92\pm0.09$ &  $0.65\pm0.07$\\	
MACS J0911.2+1746   & 1.2 & $-0.78\pm0.16$ &  $0.53\pm0.08$\\ 
MACS J2214.9-1359   & 1.3 &  $-0.76\pm0.12$ &  $0.71\pm0.09$\\ 
MACS J0257.1-2325   & 1.3 & $+0.02\pm0.31$ &  $0.35\pm0.07$\\ 
A2199  & 48 & $-1.12\pm0.06$ &  $0.48\pm0.04$\\
A1656  & 48 & $-1.02\pm0.05$ &  $0.44\pm0.04$\\
\hline                                                      
\end{tabular} 
\end{table}

\subsection{Some advantages and shortcomings of current studies}

Other authors argue that $L/F$ or its reciprocal, $F/L$, are
preferable to $\alpha$ in the study of the relative 
abundance of faint red galaxies, usually
with the rationale that the Schechter function might not
describe the luminosity function in the studied mag range.
Beside the fact that the very same authors find acceptable
fit on their data (typically, $\chi^2_v \approx 1$ values), and 
thus they argue something not supported by their own data, we note that
$L/F$, and its reciprocal, $F/L$,
both are quantities difficult to manage from a 
statistical point of view. For example an average value,
computed by a weighted sum, or a fit performed minimizing the
$\chi^2$, has a special meaning,
because the result depends on whether 
$L/F$ or $F/L$ is averaged (fitted). 
For example, let's consider two, for sake of clarity, data points, $(f/l)_1=3\pm0.9$ 
and $(f/l)_2=0.3333\pm0.1$ and two possible averages.
The error weighted average $\langle f/l \rangle$ is $0.37$. The
error weighted average $\langle l/f \rangle$ of the
reciprocal values ($(l/f)_i=1/(f/l)_i ; 0.3333\pm0.1$ and $3\pm0.9$) is
again $0.37$, fairly different from the reciprocal of $\langle f/l \rangle$,
$1/0.37=2.7$. Therefore $\langle f/l \rangle \ne 1/ \langle l/f \rangle$.
At first sight, by choosing the parametrization of the aimed quantity,
the astronomer
may chose the result he want. Furthermore, $\langle f/l \rangle$ has a value near to the point with index
$2$, $(f/l)_2$, whereas $\langle l/f \rangle$ has a value near to the other data point,
with index $1$, $(f/l)_1$, a strange situation, indeed. Similar problems
are present with two data points differing by just $1 \sigma$, 
or with small samples.
Therefore, astronomers who want to use $L/F$ or $F/L$ 
are invited first to understand what is going on in the simple
case of just two measurements
and a fit with a constant (the mentioned weighted average), and then
proceed to the case they are really interested in: a few
data points and a fit with one more degree of freedom (the
redshift or mass dependence).

\begin{figure}
\psfig{figure=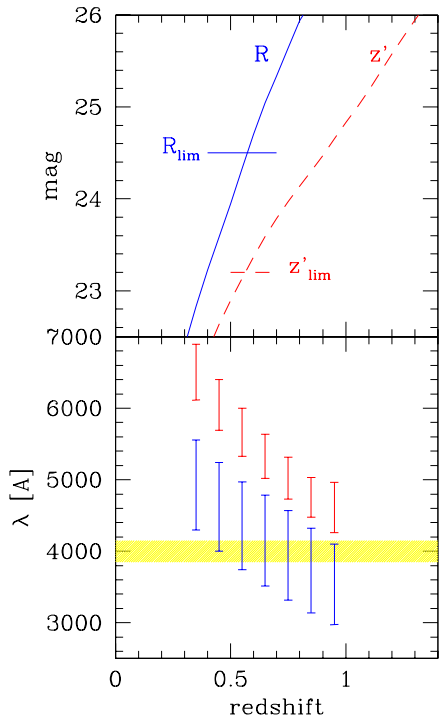,width=6truecm,clip=}
\caption[h]{{\it Bottom panel:} Rest-frame $\lambda$ sampling of the
filters used in Gilbank et al. (2007). The shaded (yellow) band marks the
4000 \AA \ break. The central wavelength of the bluest filter ($R$) 
lies longward of the Balmer break for $z < 0.75$. Compare with Fig 1.
{\it Upper panel:} $R$ and $z'$ mag of a passive evolving $M_V=-18.2$
mag galaxy (red dashed and blue solid curves, respectively) and the limiting
magnitude of the observational material used by Gilbank et al. (2007).
Gilbank et al. (2007) data are not deep enough to
detect these faint galaxies at $z\ga 0.55$.
}
\end{figure}

As mentioned,
there are currently a few determinations of the evolution of faint
red galaxies. Some of them make different claims concerning
the deficit
of faint red galaxies at high redshift, yet we have verified 
that often, but not always,  
data agree with each other, as shown in Fig. 4 and Fig. 5, 
in the (usually small) range where they are well determined.
The present work offers some advantages with respect to 
previous ones. 

-- First, our
determination is more sensitive to evolution, because
our cluster sample displays the widest redshift coverage while
keeping a large sample of clusters. 

-- Second, the present work minimizes systematics, for example using
a colour index bracketing 
the 4000 \AA \ break at every redshift. Fig 1 shows that
SDSS $u$ and $g$ filters and, at a lower extent, $B$ and $I$ filters
at $z=0.25$, sample the 4000 \AA \ break in similar way as
HST filters do at higher redshift.
This is not the rule: for example, De Lucia et al. (2007) 
use rest-frame $U-V$ at high redshift but $r'-i'$ at $z=0$ for C4
clusters ($r'$ and $i'$ have effective $\lambda=6165,7481$ \AA , respectively).
The central wavelength of the bluest filter ($R$) used in the recent 
$0.35 < z < 0.95$ study by Gilbank et al. (2007)  
lies longward of the Balmer break for $z < 0.75$ (Fig 8). 
By using the
same filter pair ($R-z'$) at all redshifts, rather than a color selection
mimicking $U-V$ at all redshifts, their technique introduces a potential bias,
as several spiral types move from red to blue between the low and the high
redshift samples because of their observational strategy. Furthermore,
at $z\ga 0.55$
the data used by Gilbank et al. (2007) are not deep enough to sample faint 
galaxies ($M_V=-18.2$ mag, such as those considered here, in Stott et al.
2007, de Lucia et al. 2007, in Gilbank \& Balogh 2008, etc.), see 
top panel of Fig 8.
In their later work, Gilbank \& Balogh (2008) use data lacking appropriate
Balmer break coverage (the $z=0.25$ point from Hansen et al. 2008 and 
the $z=0$ point from De Lucia et al. 2007) and omit data 
with better 4000 \AA \ sampling (Fig 1) and appropriate dept
(those in the present work). 
In passing,
we also disagree with their statement about the number of clusters at
low redshift in our sample: their claim that it
``only contains two $z<0.5$ clusters", while instead our
sample includes 12 clusters, 10 from
Smail et al. (1998) and two from our own analysis of SDSS data.

-- Third, interlopers are removed using observations taken in the very 
same bands as cluster observations (see Table 1), to avoid systematics
(see Smail et al. 1998 and Andreon et al. 2006). 
This is often not the case: for cluster and control field
Stott et al. (2007) use different filters, whereas de Lucia et al. (2007)
use different telescopes and
filters.
The impact of these systematics
is not quantified in the mentioned
works. 

--Fourth, we feel our statistical analysis to be preferable:
beside already mentioned statistical considerations,
there are a number of debatable issues in other works, such as 
averages of incompatible measurements, Poisson errors for
binomial distributed quantities, and unphysical results such as 
negative number of galaxies.

-- Finally, our clusters are spectroscopically
confirmed and have an X-ray emission that confirms the existence of
deep potential wells. Instead, we ignore whether candidate, or putative, 
clusters without a spectroscopic confirmation or an X-ray detection,
studied in some other papers (e.g. Kodama et al. 2004)
are clusters or line of sight superpositions.  Sometimes,
followup spectroscopic observations shows that a considerable fraction
of them are line of sight superpositions (e.g. Yamada et al. 2005).

\section{Summary and Conclusions}

The history of mass assembly of bright (massive) red galaxies in clusters
is pretty well known: 
they were assembled at early times, 
as testified by the passive evolution of 
their characteristic magnitude (e.g. de Propris et al. 1998, 2007; Andreon
et al. 2007), the constancy of 
their stellar mass function (e.g. Andreon 2006b)
and of the halo occupation
number, i.e. the number of galaxies per unit cluster mass 
(Lin et al. 2006, Andreon et al. 2008).
We stress that all mentioned works favour the above scenario, but only one,
(Andreon 2006b), excludes contender models, and we emphasize that most mentioned 
works have samples that are dominated, but not exclusively composed, 
by red galaxies.

The history of mass assembly of faint red galaxies is far less clear. 
The present paper shows that a non-evolution of
the faint end slope $\alpha$, or any related number such as 
the luminous-to-faint ratio, is fully compatible with the data. 
This implies that the history of mass assembly of faint red galaxies
is strictly parallel to the one of their massive cousins, in order
to keep the relative abundance constant. Therefore,
the build up of the red sequence is largely complete by $z=1.3$ down
to $0.05 L^*$, and,
if a differential filling is envisaged, it should occur mostly
at much larger redshift. Similarly, cluster mass, as parametrized
by X-ray luminosity or velocity dispersion, seems not to play
any role in shaping the relative abundance of faint galaxies, contrary
to some previous claims.
Our claims are based on one of the largest samples, 
spread over the wider redshift range studied thus far with
a large cluster sample, with great attention to systematics.
A recent ($z<1.3$) transformation
of many blue galaxies in faint red galaxies would 
modify the faint-end slope of the LF, change the $F/L$ ratio 
and inflate the color scatter of the colour-magnitude relation,
none of which have been observed. Yet, a redshift trend is expected 
because of the spiral morphology of some faint red galaxies in nearby clusters, 
but a larger sample of clusters (at $z \gg 0.5 $) is needed
to measure its small amplitude. The present sample is however large enough 
to discard the claimed steep trends previously suggested in literature.

\section*{Acknowledgments}

I acknowledge  the referee for helping me to improve the presentation
of this work,
Roberto De Propris, Yusei Koyama, Masayuki Tanaka, 
Tommaso Treu for useful
comments to the draft, Gabriella De Lucia for clarifications about
her results and Kevin Pimbblet, Nelson Caldwell, Duncan Forbes 
for helping me with recovering the Tarlevich et al. (2001) Coma catalog 
published electronically in MNRAS, but not available on the journal site.
 
For the standard SDSS acknowledgment see: 
http://www.sdss.org/dr5/coverage/credits.html

We thanks HST programs 9290, 9919, 9033, 9722, 9498 
9744, 9425, 9583.

\bsp

\label{lastpage}


\begin{thebibliography}{}

\bibitem[]{}
\bibitem[Adelman-McCarthy et al.(2007)]{2007ApJS..172..634A} 
Adelman-McCarthy, J.~K., et al.\ 2007, ApJS, 172, 634  

\bibitem[Andreon(1996)]{1996A&A...314..763A} 
Andreon, S.\ 1996, A\&A, 314, 763 

\bibitem[Andreon(2003)]{2003A&A...409...37A} 
Andreon, S.\ 2003, A\&A, 409, 37 

\bibitem[Andreon(2006)]{2006MNRAS.369..969A} 
Andreon, S.\ 2006a, MNRAS, 369, 969 

\bibitem[Andreon(2006)]{2006A&A...448..447A} 
Andreon, S.\ 2006b, A\&A, 448, 447 


\bibitem[Andreon et al.(2004)]{2004MNRAS.353..353A} 
Andreon, S., Willis, J., Quintana, H., Valtchanov, I., Pierre, M., \& 
Pacaud, F.\ 2004, MNRAS, 353, 353 

\bibitem[Andreon et al.(2006)]{2006MNRAS.372...60A} 
Andreon, S., Cuillandre, J.-C., Puddu, E., \& Mellier, Y.\ 2006, MNRAS, 372, 60 

\bibitem[]{}
Andreon, S. et al. 2008, MNRAS, 383, 102

\bibitem[Ball et al.(2006)]{2006MNRAS.373..845B} 
Ball, N.~M., Loveday, J., 
Brunner, R.~J., Baldry, I.~K., \& Brinkmann, J.\ 2006, MNRAS, 373, 845 

\bibitem[Balogh et al.(2004)]{2004ApJ...615L.101B} 
Balogh, M.~L., Baldry, I.~K., Nichol, R., Miller, C., Bower, R., \& 
	Glazebrook, K.\ 2004, ApJ, 615, L101 

\bibitem[Ben{\'{\i}}tez et al.(2004)]{2004ApJS..150....1B} 
Ben{\'{\i}}tez,  N., et al.\ 2004, ApJS, 150, 1 

\bibitem[Bertin \& Arnouts(1996)]{1996A&AS..117..393B} 
Bertin, E., \&  Arnouts, S.\ 1996, A\&AS, 117, 393 

\bibitem[Binggeli et al.(1988)]{1988ARA&A..26..509B} 
Binggeli, B., Sandage, A., \& Tammann, G.~A.\ 1988, ARA\&A, 26, 509 

\bibitem[Bower et al.(1992)]{1992MNRAS.254..601B} 
Bower, R.~G., Lucey, J.~R., \& Ellis, R.~S.\ 1992, MNRAS, 254, 601 

\bibitem[Bruzual \& Charlot(2003)]{2003MNRAS.344.1000B} 
Bruzual, G., \& Charlot, S.\ 2003, MNRAS, 344, 1000 

\bibitem[Butcher \& Oemler(1984)]{1984ApJ...285..426B} 
Butcher, H.~\& Oemler, A.\ 1984, ApJ, 285, 426


\bibitem[Cool et al.(2006)]{2006AJ....131..736C} 
Cool, R.~J., Eisenstein, D.~J., Johnston, D., Scranton, R., 
	Brinkmann, J., Schneider, D.~P., \& Zehavi, I.\ 2006, AJ, 131, 736 


\bibitem[]{}
D'Agostini G., 2003, {\it Bayesian reasoning in data analysis: 
A critical introduction}, World Scientific Publishing.

\bibitem[]{}
D'Agostini G., 2005, preprint (physics/0511182)


\bibitem[De Lucia et al.(2007)]{2007MNRAS.374..809D} 
De Lucia, G., et al.\ 2007, MNRAS, 374, 809 

\bibitem[de Propris et al.(1998)]{1998ApJ...503L..45D} 
de Propris, R., Eisenhardt, P.~R., Stanford, S.~A., \& 
	Dickinson, M.\ 1998, ApJ, 503, L45 

\bibitem[De Propris et al.(2007)]{2007AJ....133.2209D} 
De Propris, R., Stanford, S.~A., Eisenhardt, P.~R., Holden, 
	B.~P., \& Rosati, P.\ 2007, AJ, 133, 2209 

\bibitem[Ebeling et al.(2007)]{2007ApJ...661L..33E} 
Ebeling, H., Barrett, E., Donovan, D., Ma, C.-J., Edge, 
	A.~C., \& van Speybroeck, L.\ 2007, ApJ, 661, L33 

\bibitem[Eisenhardt et al.(2007)]{2007ApJS..169..225E} 
Eisenhardt, P.~R., De Propris, R., Gonzalez, A.~H., Stanford, S.~A., 
	Wang, M., \& Dickinson, M.\ 2007, ApJS, 169, 225 

\bibitem[Ettori et al.(2004)]{2004A&A...417...13E} 
Ettori, S., Tozzi, P., Borgani, S., \& Rosati, P.\ 2004, A\&A, 417, 13 

\bibitem[]{}
Ford, H. C., et al. 1998, Proc. SPIE, 3356, 234

\bibitem[]{}
Ford, H. C., 2002, Proc. SPIE, 4854, 81 

\bibitem[Gal \& Lubin(2004)]{2004ApJ...607L...1G} 
Gal, R.~R., \& Lubin, L.~M.\ 2004, ApJ, 607, L1 

\bibitem[Garilli et al.(1999)]{1999A&A...342..408G} 
Garilli, B., Maccagni, D., \& Andreon, S.\ 1999, A\&A, 342, 408 

\bibitem[Gilbank et al.(2007)]{2007arXiv0710.2351G} 
Gilbank, D.~G., Yee, H.~K.~C., Ellingson, E., Gladders, M.~D., Loh, Y.~-., Barrientos, L.~F., \& 
Barkhouse, W.~A.\ 2007, ApJ, in press (arXiv:0710.2351)

\bibitem[]{}
Gilbank, D.~G \& Balogh M. L., 2008, MNRAS, in press (arXiv:0801.1930)

\bibitem[]{}
Hack, W. 1999, CALACS Operation and Implementation, Instrument Science Report
ACS 99-03 (Baltimore: STScI)

\bibitem[Hansen et al.(2008)]{2007arXiv0710.3780H}
Hansen, S.~M., Sheldon, E.~S., Wechsler, R.~H.,
\& Koester, B.~P.\ 2008, ApJ, submitted (arXiv:0710.3780)

\bibitem[Hogg et al.(2004)]{2004ApJ...601L..29H} 
Hogg, D.~W., et al.\ 2004, ApJ, 601, L29 

\bibitem[Jorgensen et al.(1996)]{1996MNRAS.280..167J} 
Jorgensen, I., Franx, M., \& Kjaergaard, P.\ 1996, MNRAS, 280, 167

\bibitem[Kodama, Arimoto, Barger, \& Arag'on-Salamanca(1998)]{1998A&A...334...99K} 
Kodama, T., Arimoto, N., Barger, A.~J., \& Arag'on-Salamanca, A.\ 1998, A\&A, 33
4, 99

\bibitem[Kodama et al.(2004)]{2004MNRAS.350.1005K} 
Kodama, T., et al.\  2004, MNRAS, 350, 1005 

\bibitem[Koyama et al.(2007)]{2007arXiv0710.0632K} 
Koyama, Y., Kodama, T., Tanaka, M., Shimasaku, K., \& Okamura, S.\ 2007, MNRAS,
in press (arXiv:0710.0632) 

\bibitem[Koekemoer et al.(2002)]{2002hstc.conf..337K} 
Koekemoer, A.~M., Fruchter, A.~S., Hook, R.~N., \& Hack, W.\ 2002, in
The 2002 HST Calibration 
Workshop : Hubble after the Installation of the ACS and the NICMOS Cooling 
System, Proceedings of a Workshop 

\bibitem[Joy et al.(2001)]{2001ApJ...551L...1J} 
Joy, M., et al.\ 2001, ApJ, 551, L1 

\bibitem[Lin et al.(2006)]{2006ApJ...650L..99L} 
Lin, Y.-T., Mohr, J.~J., 
Gonzalez, A.~H., \& Stanford, S.~A.\ 2006, ApJ, 650, L99 

\bibitem[Lubin et al.(2004)]{2004ApJ...601L...9L} 
Lubin, L.~M., Mulchaey, J.~S., \& Postman, M.\ 2004, ApJ, 601, L9 

\bibitem[Maughan et al.(2004)]{2004MNRAS.351.1193M} 
Maughan, B.~J., Jones, L.~R., Ebeling, H., \& Scharf, C.\ 2004, 	
	MNRAS, 351, 1193 




\bibitem[]{} 
Oemler, A., Jr. 1992, in Clusters and Superclusters of Galaxies, 
	ed. A.C. Fabian (Dordrecht: Kluwer), 29

\bibitem[Pahre et al.(1998)]{1998AJ....116.1606P} 
Pahre, M.~A., de Carvalho, R.~R., \& Djorgovski, S.~G.\ 1998, 
	AJ, 116, 1606 

\bibitem[Pimbblet et al.(2002)]{2002MNRAS.331..333P} 
Pimbblet, K.~A., 
Smail, I., Kodama, T., Couch, W.~J., Edge, A.~C., Zabludoff, A.~I., \& 
O'Hely, E.\ 2002, MNRAS, 331, 333 

\bibitem[Schechter(1976)]{1976ApJ...203..297S} 
Schechter, P.\ 1976, ApJ, 203, 297 

\bibitem[]{} 
Sivia, D. 2006, {\it Data Analysis: A Bayesian Tutorial}, Oxford University
Press

\bibitem[Smail et al.(1998)]{1998MNRAS.293..124S} 
Smail, I., Edge, A.~C., Ellis, R.~S., \& Blandford, R.~D.\ 1998, MNRAS, 293, 124 

\bibitem[Stanford, Eisenhardt, \& Dickinson(1998)]{1998ApJ...492..461S} 
Stanford, S.~A., Eisenhardt, P.~R., \& Dickinson, M.\ 1998, ApJ, 492, 461

\bibitem[Stanford et al.(2001)]{2001ApJ...552..504S} 
Stanford, S.~A., Holden, B., Rosati, P., Tozzi, P., Borgani, S., 
	Eisenhardt, P.~R., \& Spinrad, H.\ 2001, ApJ, 552, 504 

\bibitem[Stott et al.(2007)]{2007ApJ...661...95S} 
Stott, J.~P., Smail, I., Edge, A.~C., Ebeling, H., Smith, G.~P., 
	Kneib, J.-P., \& Pimbblet, K.~A.\ 2007, ApJ, 661, 95 
	
\bibitem[]{}
Schwarz G., 1978, Annals of Statistics, 5, 461

\bibitem[Terlevich et al.(2001)]{2001MNRAS.326.1547T} 
Terlevich, A.~I., Caldwell, N., \& Bower, R.~G.\ 2001, MNRAS, 326, 1547 

\bibitem[van Dokkum \& van der Marel(2007)]{2007ApJ...655...30V} 
van  Dokkum, P.~G., \& van der Marel, R.~P.\ 2007, ApJ, 655, 30 

\bibitem[Yamada et al.(2005)]{2005ApJ...634..861Y} 
Yamada, T., et al.\ 2005, ApJ, 634, 861 



 

\end{thebibliography}
\end{document}